\documentclass[aps,prb,twocolumn,superscriptaddress,nofootinbib,longbibliography]{revtex4-2}

\usepackage[english]{babel}
\usepackage[utf8]{inputenc}
\usepackage{amssymb}
\usepackage{amsmath}
\usepackage{color}
\usepackage{hyperref}
\usepackage{bbm}
\usepackage{epsfig}
\usepackage{multirow}
\usepackage[normalem]{ulem}
\usepackage[dvipsnames]{xcolor}
\usepackage{xspace}

\begin{document}

\title{Electron-phonon interaction and  electronic correlations in transport through electrostatically and tunnel coupled quantum dots}

\author{Damian Krychowski}
\email{krychowski@ifmpan.poznan.pl}
\affiliation{Institute of Molecular Physics, Polish Academy of Sciences,
			 Smoluchowskiego 17, 60-179 Pozna{\'n}, Poland}

\author{Stanis\l{}aw Lipi\'{n}ski}
\email{lipinski@ifmpan.poznan.pl}
\affiliation{Institute of Molecular Physics, Polish Academy of Sciences,
			 Smoluchowskiego 17, 60-179 Pozna{\'n}, Poland}

\date{\today}

\begin{abstract}
We investigate two equivalent capacitively and tunnel coupled quantum dots, each coupled to its own pair of leads. Local Holstein type  electron-phonon coupling  at the dots  is assumed. To study many-body effects we use the finite-U mean-field slave boson approach. For vanishing interdot interaction, weak e-ph coupling and finite tunneling, molecular orbital spin Kondo effects occur for single electron or single hole occupations. Phonons influence both correlations and tunneling and additionally they shift the energies of the dots. Depending on the dot energies and the strength of electron-phonon coupling, the system is occupied by a different number of electrons that effectively interact with each other repulsively or attractively leading to a number of different ground states of DQD. Among them are  Kondo-like states with  spin, orbital or charge correlations resulting from polaron cotunneling processes and states with magnetic intersite correlations.
\end{abstract}


\maketitle

\section{Introduction}
In double quantum dots (DQD) both spin and orbital degrees of freedom are relevant, which leads to creating various correlated states. Multi-dot devices allow studying these phenomena  in more detail than in bulk systems, because  at nano scale it is possible to tune  the couplings at will. The interest in quantum dot arrangements stems also from their potential applications for spintronics and quantum computation \cite{Burkard,Awschalom}. Depending on DQD geometry (series or parallel coupled) continuous or discontinuous transitions between different ground states have been predicted \cite{Lopez2002}. Most of the papers concern the dots connected in series, but recently, it has been realized that parallel double quantum dots  are much more experimentally suitable than dots in  series for studying spin entangled states composed of coherent Kondo resonances \cite{Chen}. Many publications, both theoretical and experimental have been already devoted to this subject (e.g. \cite{Martins1,Martins2,Choi,Vernek,Krychowski,Keller2014}). Recently nano-electromechanical systems (NEMS) have attracted much interest, because they integrate electrical and mechanical functionalities \cite{Leturcq, Blick}. Particularly attractive in this respect are molecular systems due to their softness. Among the hot topics studied is  the impact of electron - phonon (e-ph) coupling on electron correlations \cite{Mravlje,Florkow,Cornaglia1,Rudzinski,Wang2012,Cornaglia2,Bocian2015}. A very useful method of experimental study of these subtle effects are the conductance measurements e.g. observation of  Kondo phonon satellite peaks  \cite{Rakhmilevitch}.  Similar phenomena have been also reported in the rigid structures of semiconductor QDs  embedded in a freestanding GaAs/AlGaAs membrane \cite{Weig}. Advanced nanotechnology has been able to provide a good morphology manipulation of semiconductor QDs such as size, shape  strain distribution, and inhomogeneities, which determine  the effective e- ph interaction \cite{Kuo}. Similar  research opportunities for  studying impact of phonons on transport for  different phonon frequencies and different coupling constants exist in carbon nanotubes, since  these quantities depend on the microscopic details of CNT,  such as its chirality and radius \cite{Popov,Mariani}.

In this paper, we investigate the role of e-ph interaction in the occurrence of various highly correlated DQD behaviors and the reflection of these phenomena in transport properties. We examine impact of phonons on the  competition between Kondo phases and local singlet phases in spin and charge degrees of freedom. If the vibration energy exceeds the hybridization energy between dot and electrode electrons (antiadiabatic limit), then strong local Holstein e-ph coupling drives the system towards polaron states localized at the single dot. E-ph coupling introduces attractive interdot interaction and suppresses tunneling. For intermediate e-ph coupling, when phonon induced attractive interaction is smaller than Coulomb repulsion and tunneling to the leads is not completely suppressed by phonons. For weak and intermediate interdot tunneling  and weak e-ph coupling  single site  spin Kondo effects occur for odd occupancies of DQD and they transform into molecular spin Kondo states for stronger interactions with phonons.  For still higher values of  e-ph coupling transitions into higher occupancies result and Kondo resonances are destroyed. When two electrons reside on the dots and e-ph coupling is weak  local spin singlet forms.  For stronger interactions with phonons subsequent transitions into different Kondo states occur. For the electrostatically decoupled dots successive states appear with increasing e-ph interactions:  orbital Kondo, two-site Kondo and charge-orbital Kondo states. The latter occurs when phonon induced attraction exceeds Coulomb repulsion. For electrostatically coupled dots the similar sequence of states for increasing e-ph coupling in double occupancy is: orbital Kondo,  two- site Kondo state and isospin correlated Kondo state.

\section{Model and formalism}
We consider two equivalent single-level  coupled  quantum dots, each coupled to its own pair of leads (Fig.1).  The local electron densities at the dots are coupled to local vibrations.  The system is modeled by Anderson-Holstein two-impurity Hamiltonian:
\begin{eqnarray}
&&H=H^{e}+H^{ph}+H^{e-ph}
\end{eqnarray}
where
\begin{eqnarray}
&&\nonumber H^{e}=\sum_{ls}E_{d}n_{ls}+\sum_{s}t(d^{\dagger}_{1s}d_{2s}+h.c.)+U\sum_{l}n_{l\uparrow}n_{l\downarrow}\\
&&\nonumber+U'\sum_{ss'}n_{1s}n_{2s'}+\sum_{k\alpha ls}E_{k\alpha ls}n_{k\alpha ls}+
\sum_{k\alpha ls}V(c^{\dagger}_{k\alpha ls}d_{ls}+h.c)
\end{eqnarray}
The first term represents site energies of the dots, the next tunnel coupling between the dots ($t$), the terms parameterized by $U$ and $U'$ describe intra- and interdot Coulomb interactions respectively  and the last two terms describe electrons in the electrodes and their tunneling to the dots ($V$). We assume the coupling strength to the electrodes with the rectangular density of states $1/2D$ for $|E|<D$ is $\Gamma=\pi V^{2}/2D$, with  $D$ denoting  the electron bandwidth of electrodes.
\begin{figure}
\includegraphics[width=0.88\linewidth]{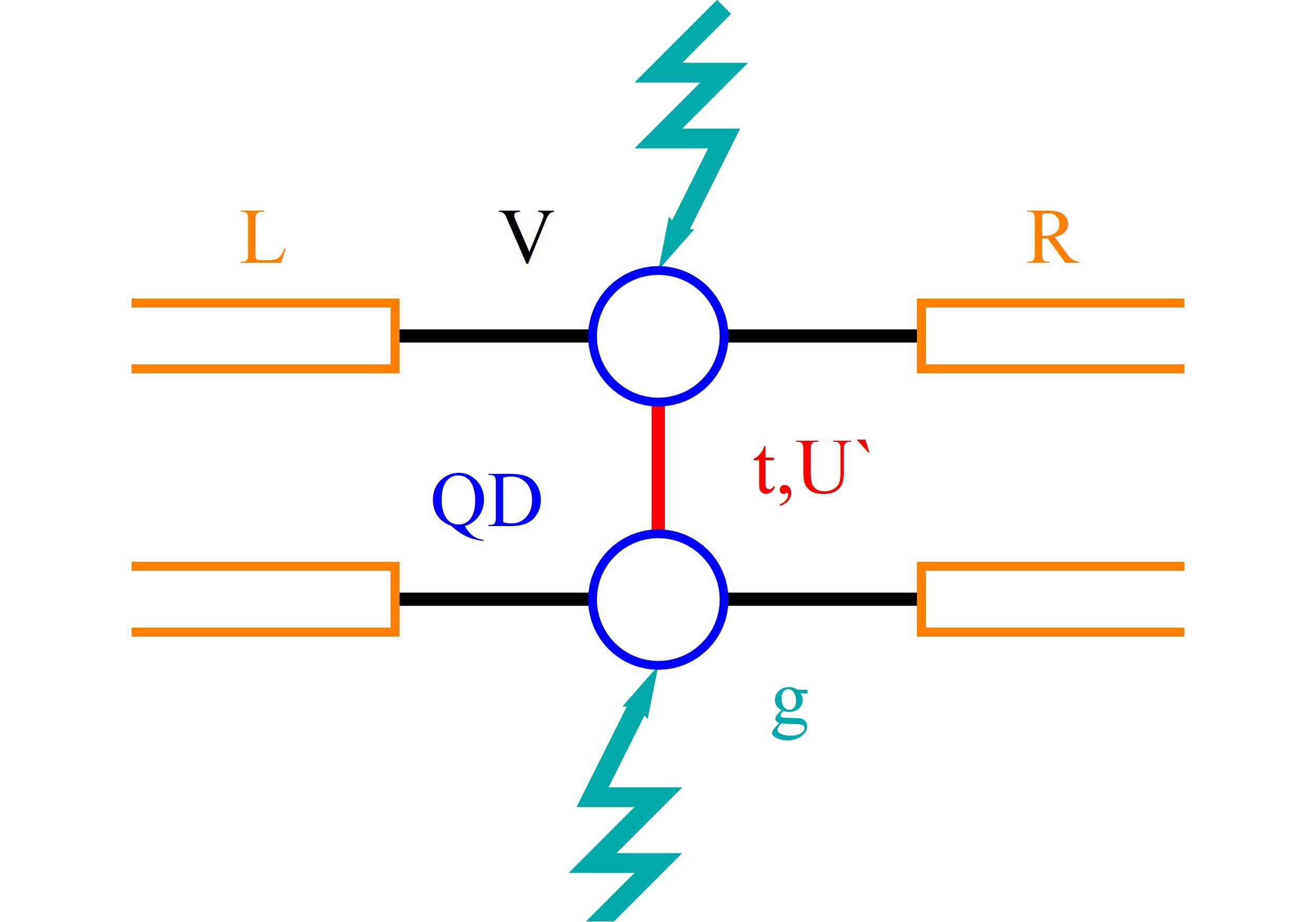}
\caption{\label{fig1} (Color online) Schematic of the parallel double-quantum dot setup attached to separate leads with interobital Coulomb interaction ($U'$)  and tunnel coupling between the dots ($t$). Hopping parameter to the electrodes is denoted by $V$. Electrons on the dots are coupled to local phonons with e-ph strength $g$ (wavy lines).}
\end{figure}

The phonon and e-ph coupling terms are given by:
\begin{eqnarray}
&&H^{ph}+H^{e-ph}=\sum_{l}\omega_{0}a^{\dagger}_{l}a_{l}+\sum_{l}g\cdot n_{l}(a^{\dagger}_{l}+a_{l})
\end{eqnarray}
The first term describes local Einstein phonons of energy  $\omega_{0}$ and the second their liner coupling to electrons at the dots.
We assume non-adiabatic regime ($\omega_{0}/\Gamma>1$)  and strong electron-phonon coupling limit, in which case one can eliminate linear e- ph coupling terms by the Lang-Firsov  unitary transformation (L-F) \cite{Lang} $\overline{H}=e^{i(S_{1}+S_{2})}He^{-i(S_{1}+S_{2})}$, defined by generators $S_{l}=(g/\omega_{0})n_{l}(a^{\dagger}_{l}-a_{l})$, ($n_{l}=\sum_{s}n_{ls}$). L-F approach is exact if $V=0$ or $g\rightarrow+\infty$. L-F transformation shifts the dots to the new equilibrium positions and in general changes the phonon vacuum. The new fermion (polaron) operators are  $\overline{d}_{ls}=d_{ls}X_{l}$ and $\overline{d}^{\dagger}_{ls}=d^{\dagger}_{ls}X^{\dagger}_{l}$ with $X_{l}=e^{-(g/\omega_{0})(a^{\dagger}_{l}-a_{l})}$. The relevant parameters of (1) become renormalized:  $\widetilde{E}_{d}=E_{d}-(g^{2}/\omega_{0})$, $\widetilde{U}=U-2(g^{2}/\omega_{0})$, $\widetilde{V}=VX_{l}$ and $\widetilde{t}=tX_{1}X_{2}$. Factor $X$ describes the effect of phonon cloud accompanying the tunneling. In the following we replace $X_{l}$ by its expectation value $\langle X\rangle =e^{-(g/\omega_{0})^{2}(n_{0}+1/2)}$ \cite{Mahan}. The electron and phonon subsystems become decoupled. Holstein coupling lowers the energy of doubly occupied orbitals with respect to singly occupied or empty ones.
It is convenient  to rewrite the effective polaron Hamiltonian in the basis of bonding ($-$) and antibonding ($+$) orbital operators $d_{\nu s}=(1/\sqrt{2})(d_{1s}\pm d_{2s})$ ($\nu=\pm$). The effective electron (polaron) Hamiltonian then reads:
\begin{eqnarray}
&&\nonumber \widetilde{H}=\sum_{\nu s}(E_{d}\pm \widetilde{t})n_{\nu s}
+\sum_{\nu}\frac{\widetilde{U}+U'}{2}(n_{\nu\uparrow}n_{\nu\downarrow}+n_{\nu\uparrow}n_{\overline{\nu}\downarrow})
\\&&+U'\sum_{s}n_{+s}n_{-s}+\frac{\widetilde{U}-U'}{2}(C_{flip}-S_{flip})
\\&&\nonumber+\sum_{k\alpha \nu s}E_{k\alpha \nu s}n_{k\alpha \nu s}+
\sum_{k\alpha \nu s}\widetilde{V}(c^{\dagger}_{k\alpha \nu s}d_{\nu s}+h.c)
\end{eqnarray}
The last term represents charge-flip $C_{flip}=T^{+}_{+}T^{-}_{-}+h.c.$ and spin-flip $S_{flip}=S^{+}_{+}S^{-}_{-}+h.c.$ processes, where spin and orbital isospin are given by the fermionic operators as follows:
\begin{eqnarray}
&&\nonumber S^{Z}_{\nu}=(n_{\nu\uparrow}-n_{\nu\downarrow})/2
\\&&S^{-}_{\nu}=d^{\dagger}_{\nu\downarrow}d_{\nu\uparrow}=(S^{+}_{\nu})^{\dagger}
\end{eqnarray}
and
\begin{eqnarray}
&&\nonumber T^{Z}_{\nu}=(n_{\nu}-1)/2
\\&&T^{-}_{\nu}=d_{\nu\uparrow}d_{\nu\downarrow}=(T^{+}_{\nu})^{\dagger}.
\end{eqnarray}

To discuss correlations we use generalized finite -U slave boson mean field approximation (SBMFA) of Kotliar and Ruckenstein \cite{Kotliar,Dong2,Krychowski2}. Compared to the  more exact numerical approaches e.g. most powerful method - renormalization group approach (NRG) \cite{Bulla},  SBMFA has narrower range of applicability.  Whereas NRG can treat systems with a broad and continuous spectrum of energies for arbitrary temperatures,  SBMFA is correct only in the unitary Kondo regime or close to it, but also leads to local Fermi-liquid behavior at zero temperature. It gives reliable results of linear conductance also for systems with weakly broken symmetry, as confirmed by experiments and NRG calculations \cite{Manteli,Souza}. SBMFA  breaks down at higher temperatures, but going beyond mean field by taking into account slave boson  fluctuations would remove this deficiency. SBMFA  has higher computational efficiency than  NRG and allows analysis  of nonlinear transport. In the context of e- ph problem considered in the following,  it is worth mentioning,  that only a limited number of bosonic states can be taken into account in numerical diagonalization carried out within  NRG and  this imposes some restrictions on phonon frequencies and e- ph coupling strengths that may be considered in NRG procedure \cite{Bulla}. For analysis of  correlation effects we  introduce, in the spirit of SB approach, a set of boson operators for each electronic configuration of the system. The auxiliary bosons $e, p, d, t, f$ project onto empty, single, double, triple and fully occupied states.  The single occupation projectors $p_{\nu s}$   are labeled by orbital and spin numbers, the triple occupancy bosons $t_{\nu s}$   by orbital and spin of the hole  and among d operators there are two classes $d_{\nu}$  which correspond to the occupation of the single orbital by two electrons $(\uparrow\downarrow,0), (0,\uparrow\downarrow)$ and four $d_{ss'}$ operators  representing states $(\uparrow,\uparrow), (\uparrow,\downarrow), (\downarrow,\uparrow), (\downarrow,\downarrow)$. In order to eliminate additional unphysical states introduced by SB representation one supplements SB Hamiltonian by conditions of charge conservation and completeness relations:
\begin{eqnarray}
&&\nonumber Q_{\nu s}=p^{\dagger}_{\nu s}p_{\nu s}+d^{\dagger}_{ss}d_{ss}+d^{\dagger}_{\nu}d_{\nu}+d^{\dagger}_{s-s}d_{s-s}+t^{\dagger}_{\nu s}t_{\nu s}
\\&&+\sum_{s}t^{\dagger}_{-\nu s}t_{-\nu s}+f^{\dagger}f
\end{eqnarray}
and
\begin{eqnarray}
&&\nonumber I=e^{\dagger}e+\sum_{\nu s}p^{\dagger}_{\nu s}p_{\nu s}+\sum_{\nu}d^{\dagger}_{\nu}d_{\nu}+\sum_{\nu s}d^{\dagger}_{ss'}d_{ss'}
\\&&+\sum_{\nu s}t^{\dagger}_{\nu s}t_{\nu s}+f^{\dagger}f.
\end{eqnarray}
\begin{figure}[h]
\includegraphics[width=0.8\linewidth]{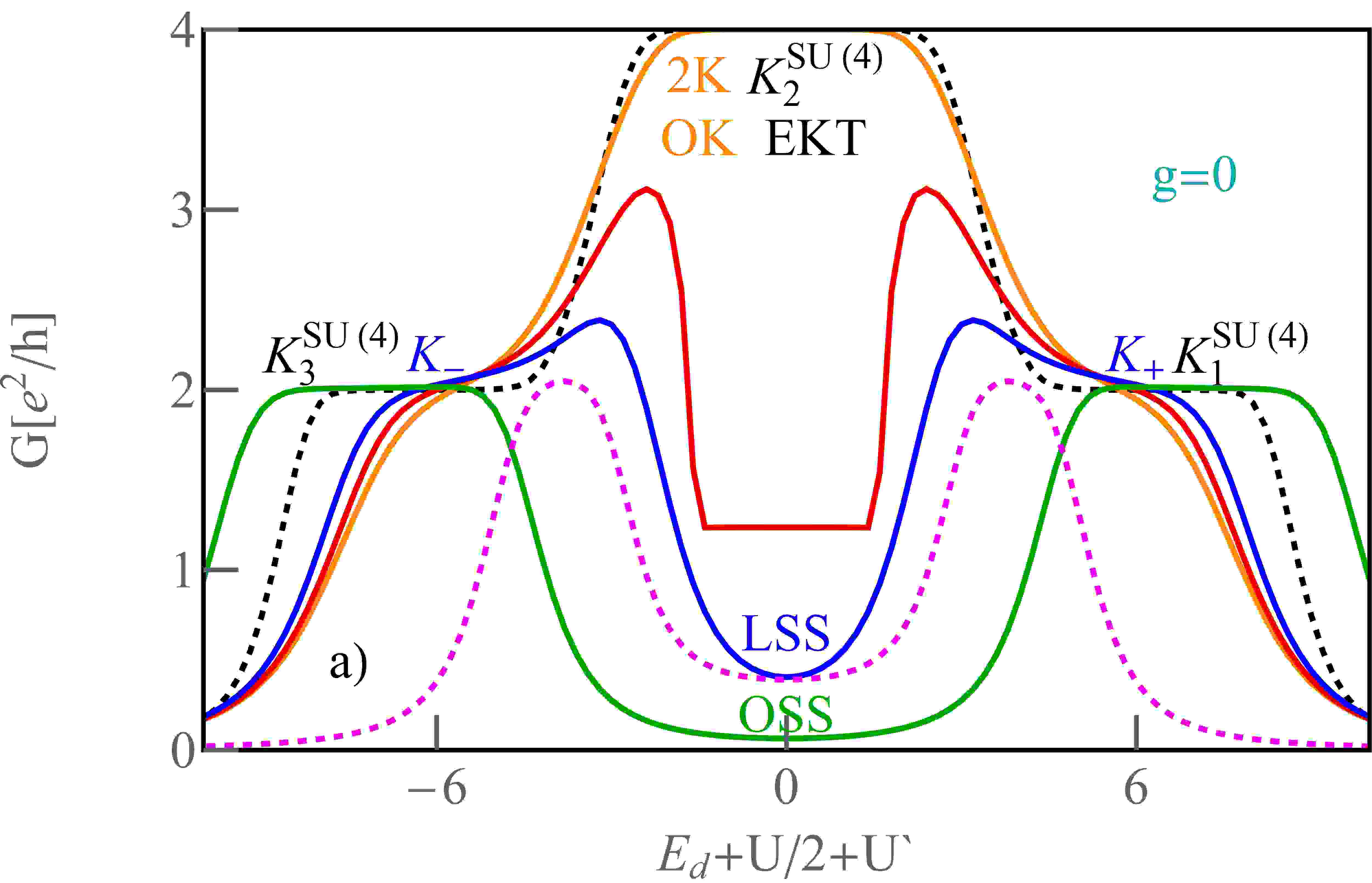}
\includegraphics[width=0.8\linewidth]{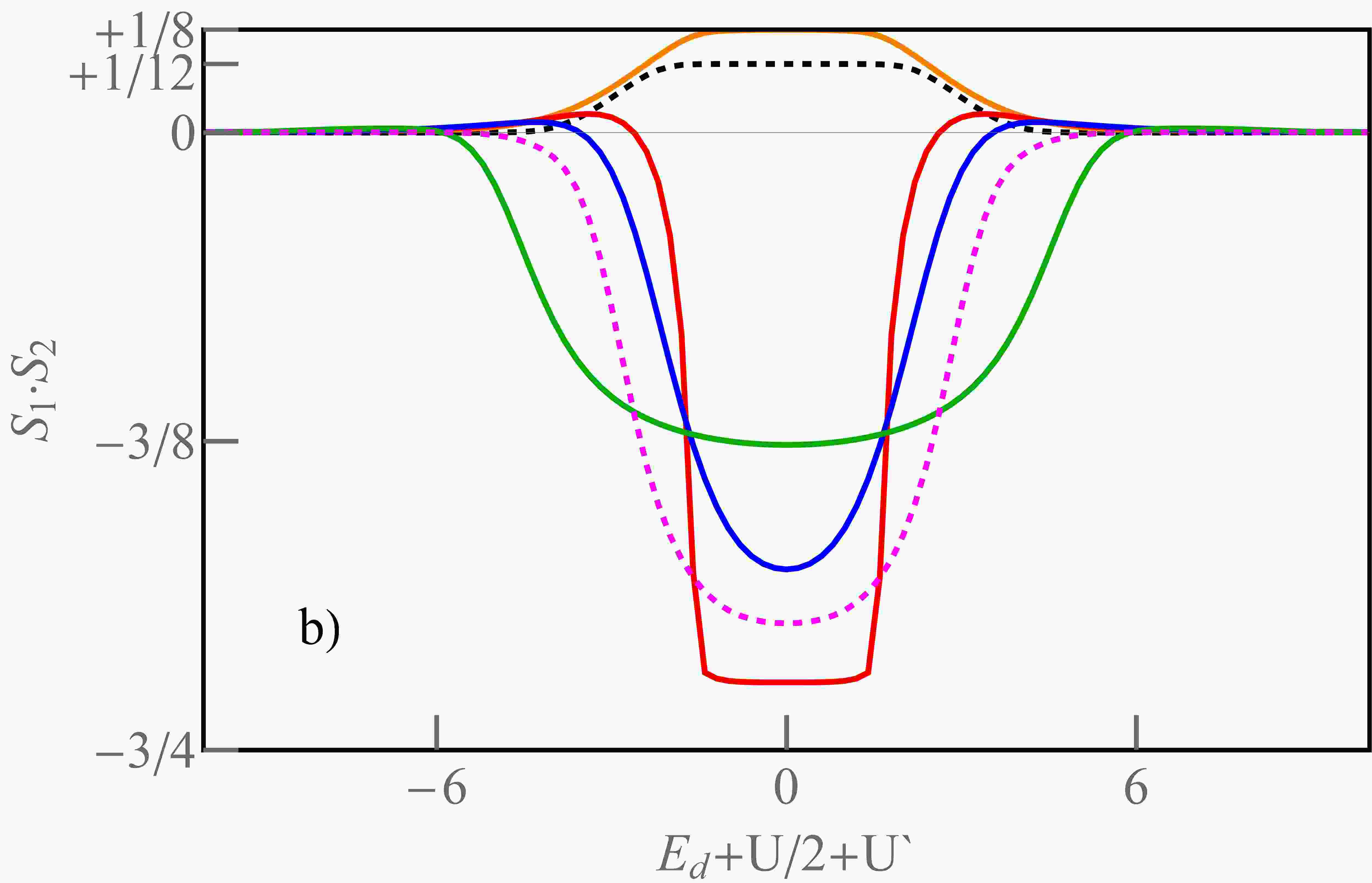}
\includegraphics[width=0.8\linewidth]{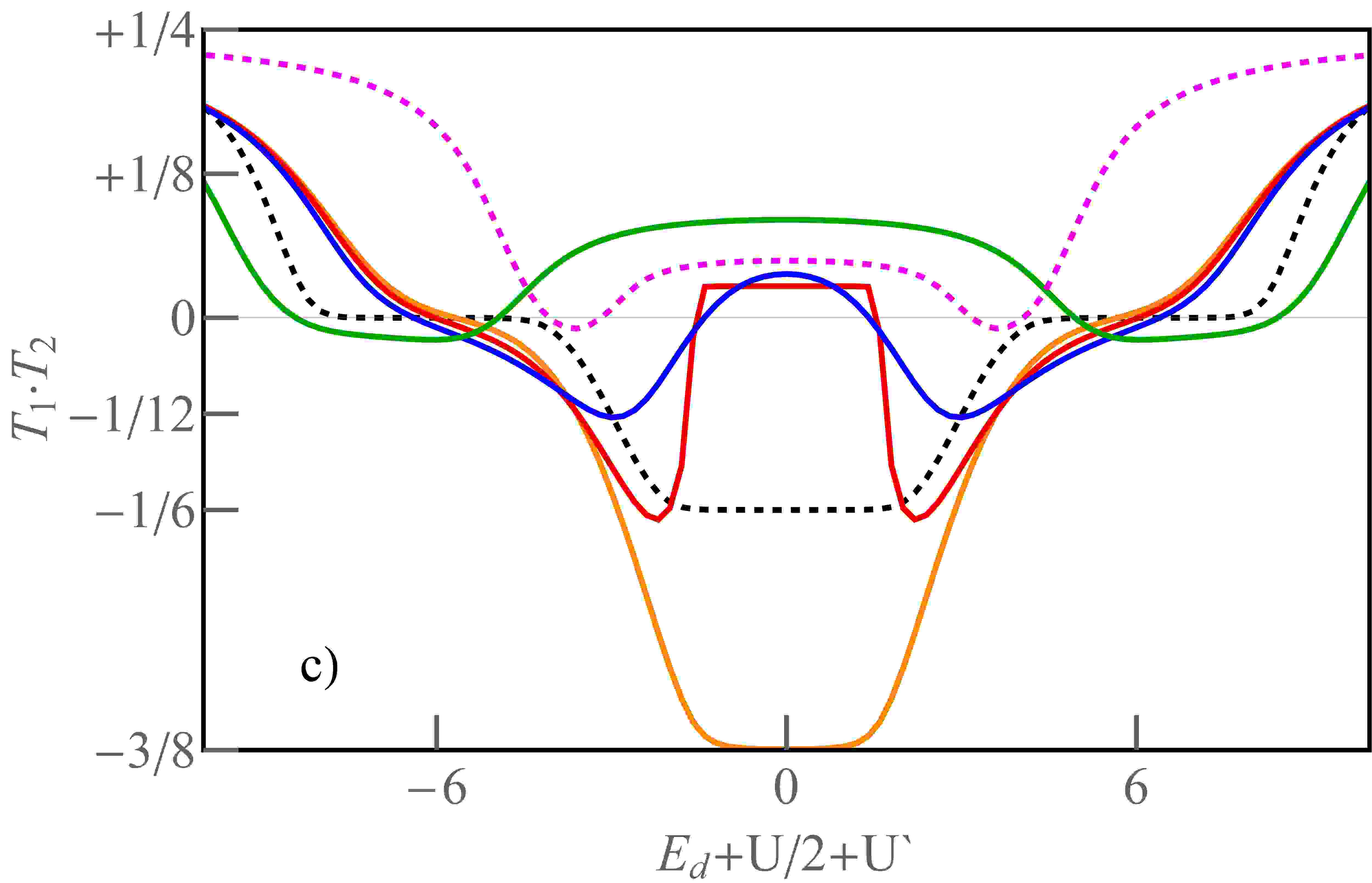}
\caption{\label{fig2} (Color online) a) Conductance $G$ as a function of dot energy level $E_{d}$ for $U=U'=6$ ($t=0$, black dashed line), $U=6, U'=5$ ($t=0, 0.2, 0.4, 2$: orange, red, blue and green curves) and $U=6, U'=0$ ($t=2$, magenta dashed line). b, c) Corresponding spin-spin $S_{1}\cdot S_{2}$ and isospin-isospin $T_{1}\cdot T_{2}$ correlators ($g=0, \Gamma=0.1$).}
\end{figure}

The constraints are incorporated into (4) via the  Lagrange multipliers $\lambda$ and $\lambda_{\nu s}$.  The corresponding K-R Hamiltonian then reads:
\begin{eqnarray}
&&\nonumber \widetilde{H}^{sbMFA}=\sum_{\nu s}(E_{d}\pm\widetilde{t})N^{f}_{\nu s}+
U'\sum_{s}d^{\dagger}_{ss}d_{ss}+
\\&&\nonumber\frac{\widetilde{U}+U'}{2}(d^{\dagger}_{s\overline{s}}d_{s\overline{s}}+\sum_{\nu}d^{\dagger}_{\nu}d_{\nu})
\\&&\frac{\widetilde{U}-U'}{2}(d^{\dagger}_{\nu}d_{\overline{\nu}}-d^{\dagger}_{s\overline{s}}d_{\overline{s}s}+h.c.)
\\&&\nonumber+\sum_{\nu s}(\widetilde{U}+2U')t^{\dagger}_{\nu s}t_{\nu s}+(2\widetilde{U}+4U')f^{\dagger}_{\nu}f_{\nu}
+\\&&\nonumber\sum_{\nu s}\lambda_{\nu s}(N_{\nu s}-Q_{\nu s})
+\lambda(I-1)+\sum_{k\alpha \nu s}E_{k\alpha \nu s}n_{k\alpha \nu s}+
\\&&\nonumber\sum_{k\alpha \nu s}\widetilde{V}(c^{\dagger}_{k\alpha \nu s}z_{\nu s}f_{\nu s}+h.c.)
\end{eqnarray}
where $N^{f}_{\nu s}=f^{\dag}_{\nu s}f_{\nu s}$ is the pseudofermion occupation number operator \cite{Kotliar}. $z_{\nu s}=(e^{\dagger}p_{\nu s}+p^{\dagger}_{\nu-s}d_{\nu}+p^{\dagger}_{-\nu s}d_{ss}+p^{\dagger}_{-\nu -s}d_{s-s}+d^{\dagger}_{-s-s}t_{-\nu-s}+d^{\dagger}_{-\nu}t_{\nu s}+d^{\dagger}_{-ss}t_{-\nu s}+t^{\dagger}_{\nu-s})/\sqrt{(1-Q_{\nu s})Q_{\nu s}}$ and $\lambda,\lambda_{\nu s}$ are the renormalization parameter and the Lagrange multiplieyers.
\begin{table*}
  \caption{Dominant slave bosons, charge fluctuations on local dot orbitals and molecular DQD orbitals, spin-spin and isospin-isospin correlators and conductance for the discussed quantum phases.}
  \label{tab:expcond}
  \begin{tabular}{lcccccc}
    \hline
    $$ & SB amplitdues & $\Delta N^{2}_{1,2}$ & $\Delta N^{2}_{\pm}$ & $\langle S_{1}\cdot S_{2}\rangle$ & $\langle T_{1}\cdot T_{2}\rangle$ & $G[e^{2}/h]$ \\
    \hline
    $K^{SU(4)}_{2}$ & $d^{2}_{\pm}=d^{2}_{ss'}=1/6$ & 0 & $1/3$ & 0 & -1/12 & 4\\
    $K^{SU(4)}_{1,3}$ & $p^{2}_{\nu s}=1/4,t^{2}_{\nu s}=1/4$ & 0 & 1/4 & 0 & 0 & 2\\
    $EKT$ & $d^{2}_{ss'}=1/4$ & 1/4 & 1/3 & $0+1/12=+1/12$ & $0-1/6=-1/6$ & 4\\
    $2K$ & $d^{2}_{ss'}=1/4$ & 1/2 & 0 & $0+1/8=+1/8$ & $-1/4-1/8=-3/8$ & 4\\
    $K_{\pm}$ & $p^{2}_{\nu s}=1/2,t^{2}_{\nu s}=1/2$ & 1/4 & 0 & 0 & 0 & 2\\
    $LSS$ & $d^{2}_{-}\approx1$ & 0 & 1 & $-1/2-1/4=-3/4$ & 0 & 0\\
    $SOK$ & $d^{2}_{\pm}=1/2$ & 0 & 1 & $-1/2-1/4=-3/4$ & 0 & 4\\
    $OSS$ & $d^{2}_{-}=1$ & 1/2 & 0 & $-1/4-1/8=-3/8$ & $1/4-1/8=+1/8$ & 0\\
    $COK$ & $e^{2}=f^{2}=d^{2}_{\pm}=1/4$ & 1 & 1 & 0 & $0+1/4=+1/4$ & 4\\
    $IOK$ & $d^{2}_{\pm}=1/2$ & 1 & 1 & 0 & $1/2-1/4=+1/4$ & 4\\
    \hline
\end{tabular}
\end{table*}
\begin{figure}[t!][h]
\includegraphics[width=0.8\linewidth]{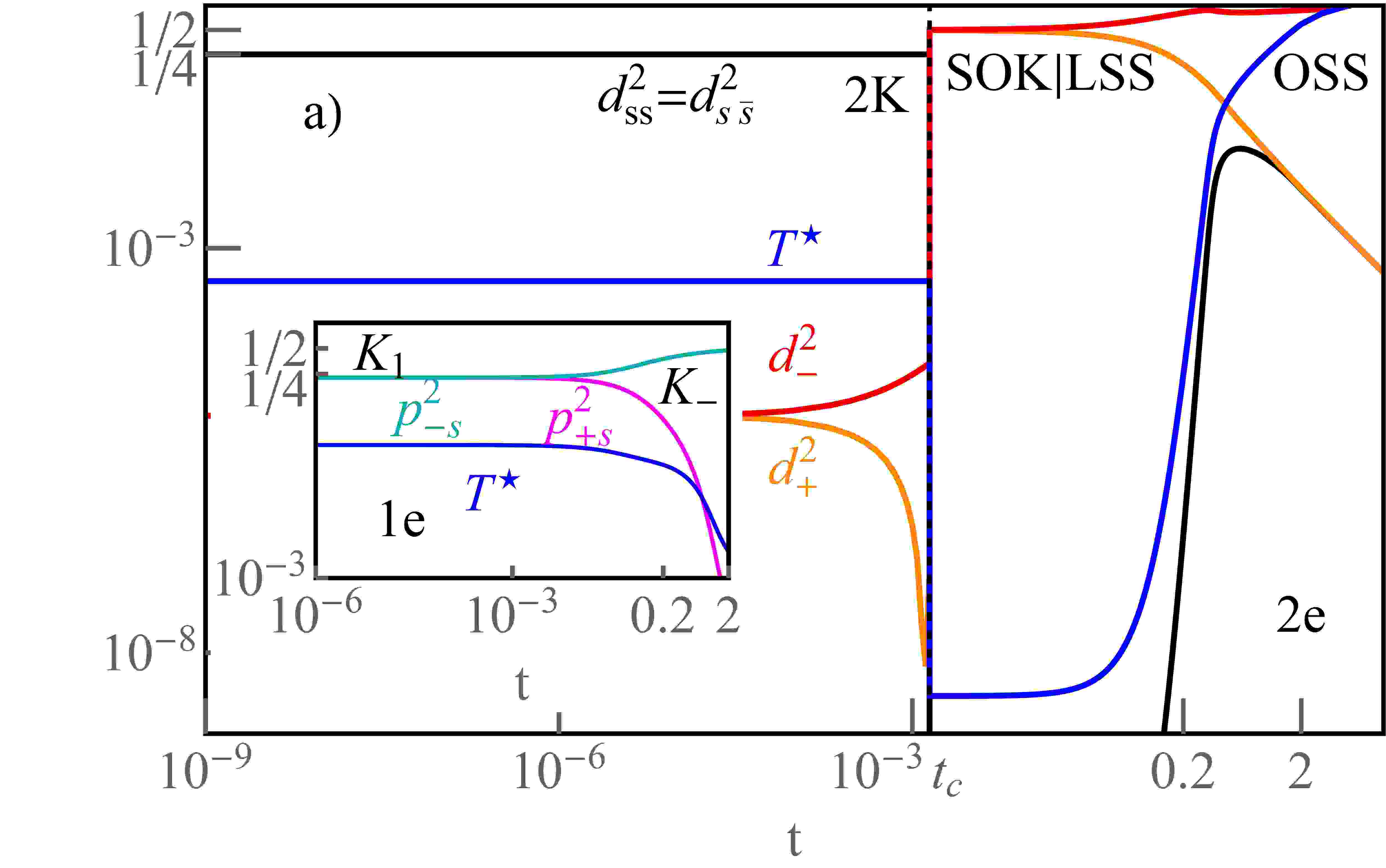}
\includegraphics[width=0.8\linewidth]{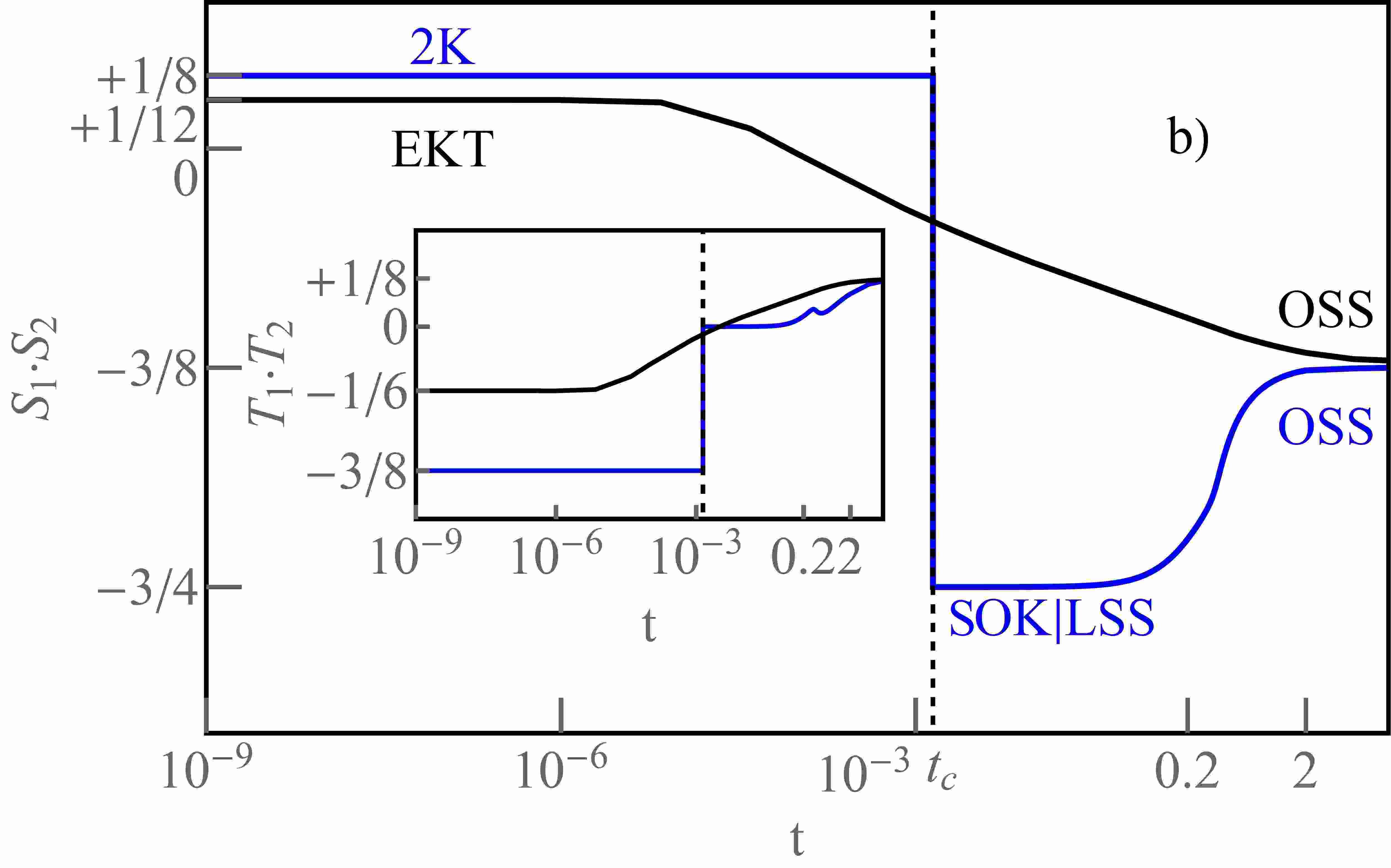}\\
\includegraphics[width=0.8\linewidth]{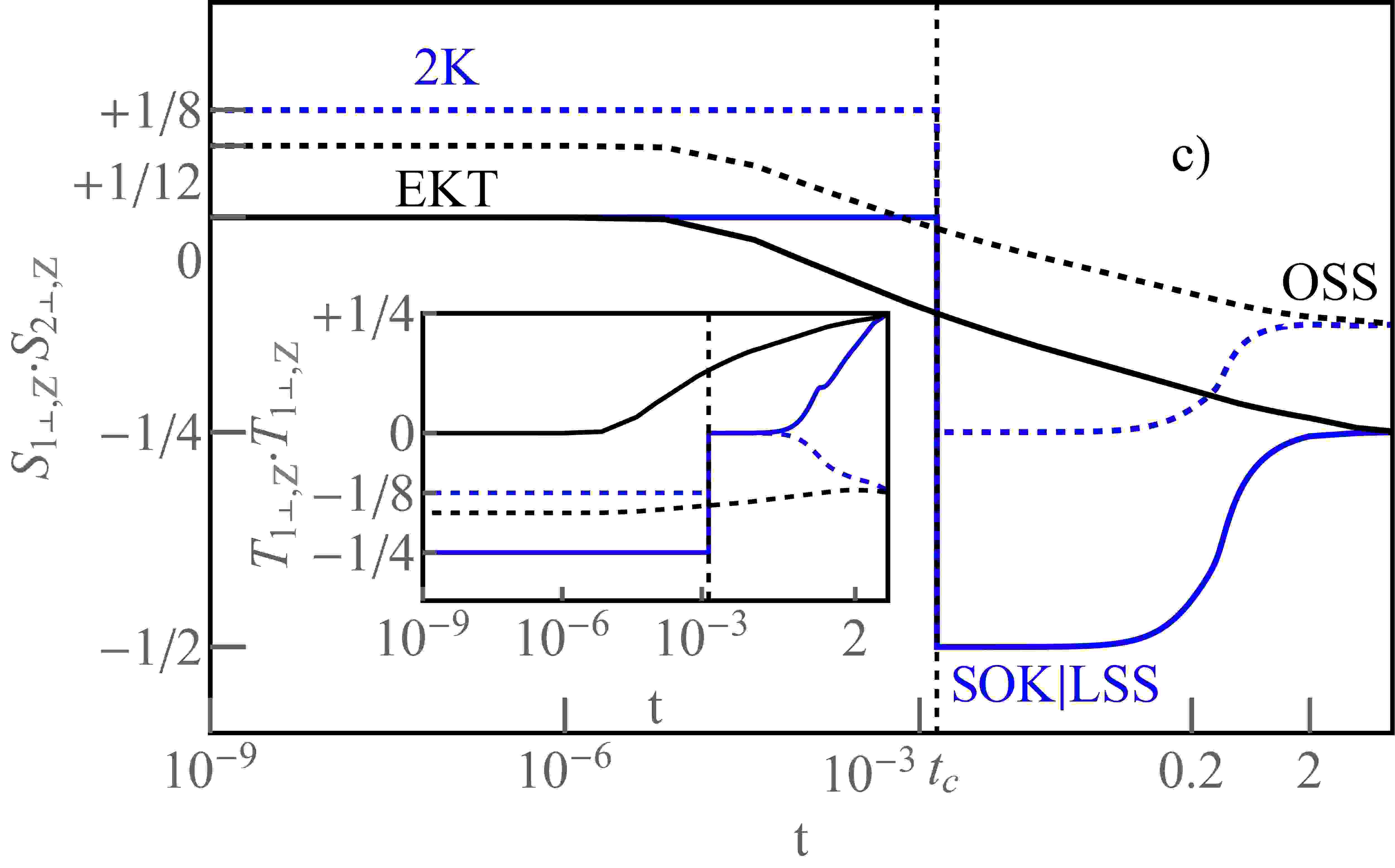}
\includegraphics[width=0.8\linewidth]{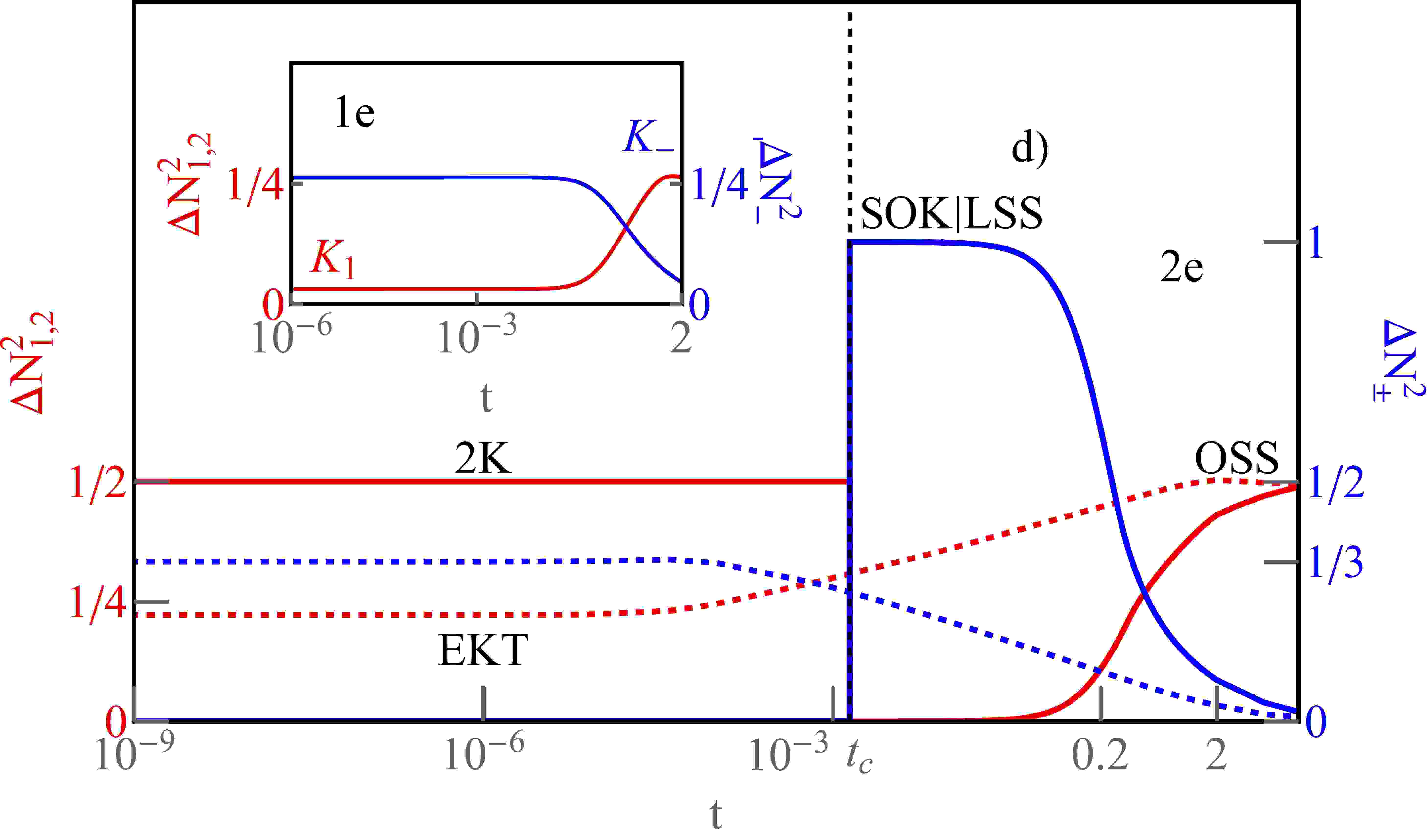}
\caption{\label{fig3} (Color online) a) Slave boson amplitudes and characteristic temperature $T^{\star}$ (blue line) ($U=6, U'=5$) for $2e$ and in 1e regime (inset). b) Total correlators $S_{1}S_{2}$ and $T_{1}T_{2}$ (inset) for $U=U'=6$ (black curves)  and $U=6 U'=5$ (blue lines).
c) Transversal($\perp$, solid lines) and longitudinal($z$, dashed lines) parts of spin-spin and isospin-isospin correlators
for $U=U`=6$ (black curves) and $U=6 U`=5$ (blue lines). d) Orbital $\Delta N^{2}_{\pm}$ and dot $\Delta N^{2}_{1,2}$ fluctuations
for $2e$ region (solid and dashed lines are for $U=U'=6$ and $U=6, U'=5$ respectively) and charge fluctuations for $1e$ (inset).}
\end{figure}
\begin{figure}
\includegraphics[width=0.8\linewidth]{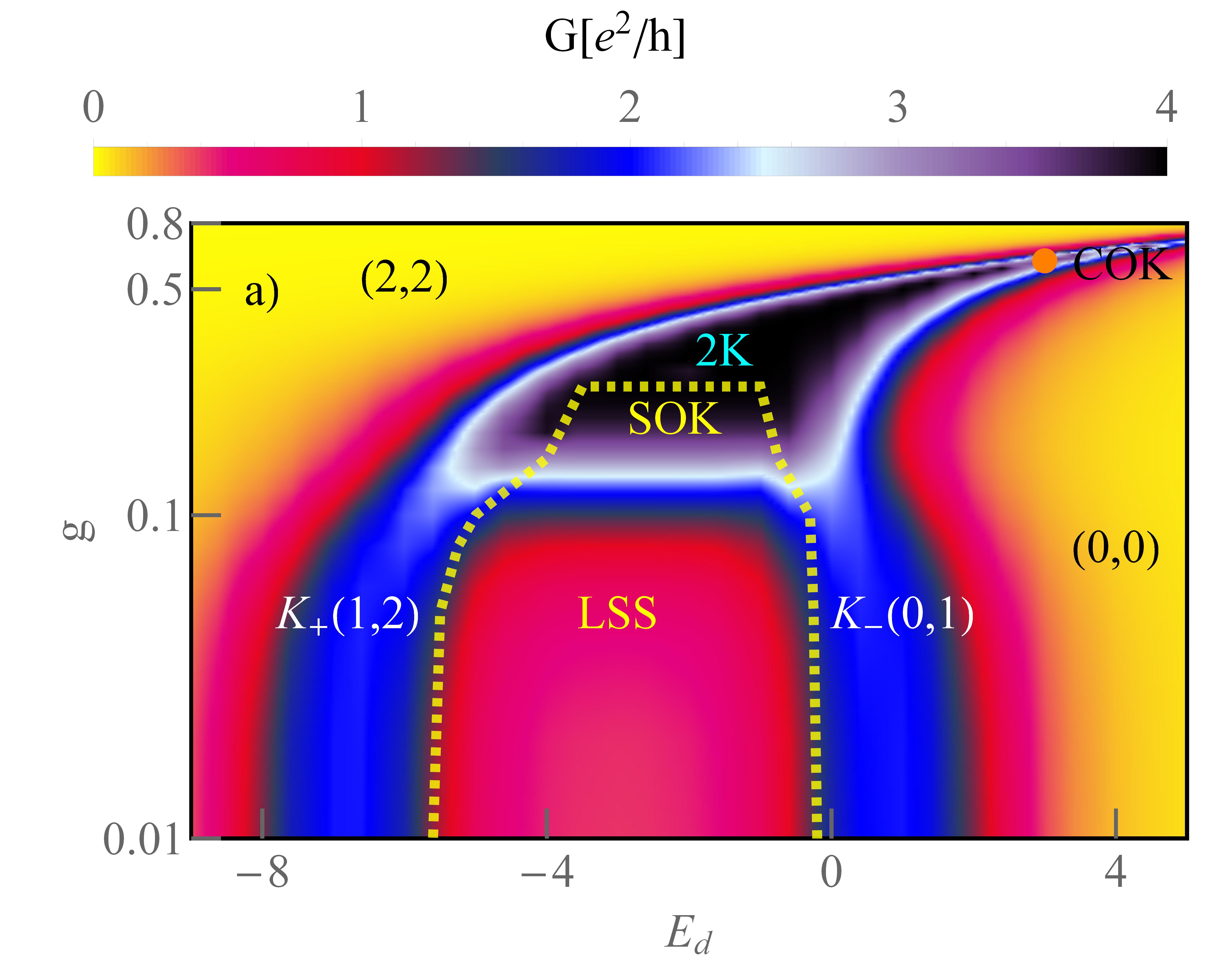}
\includegraphics[width=0.8\linewidth]{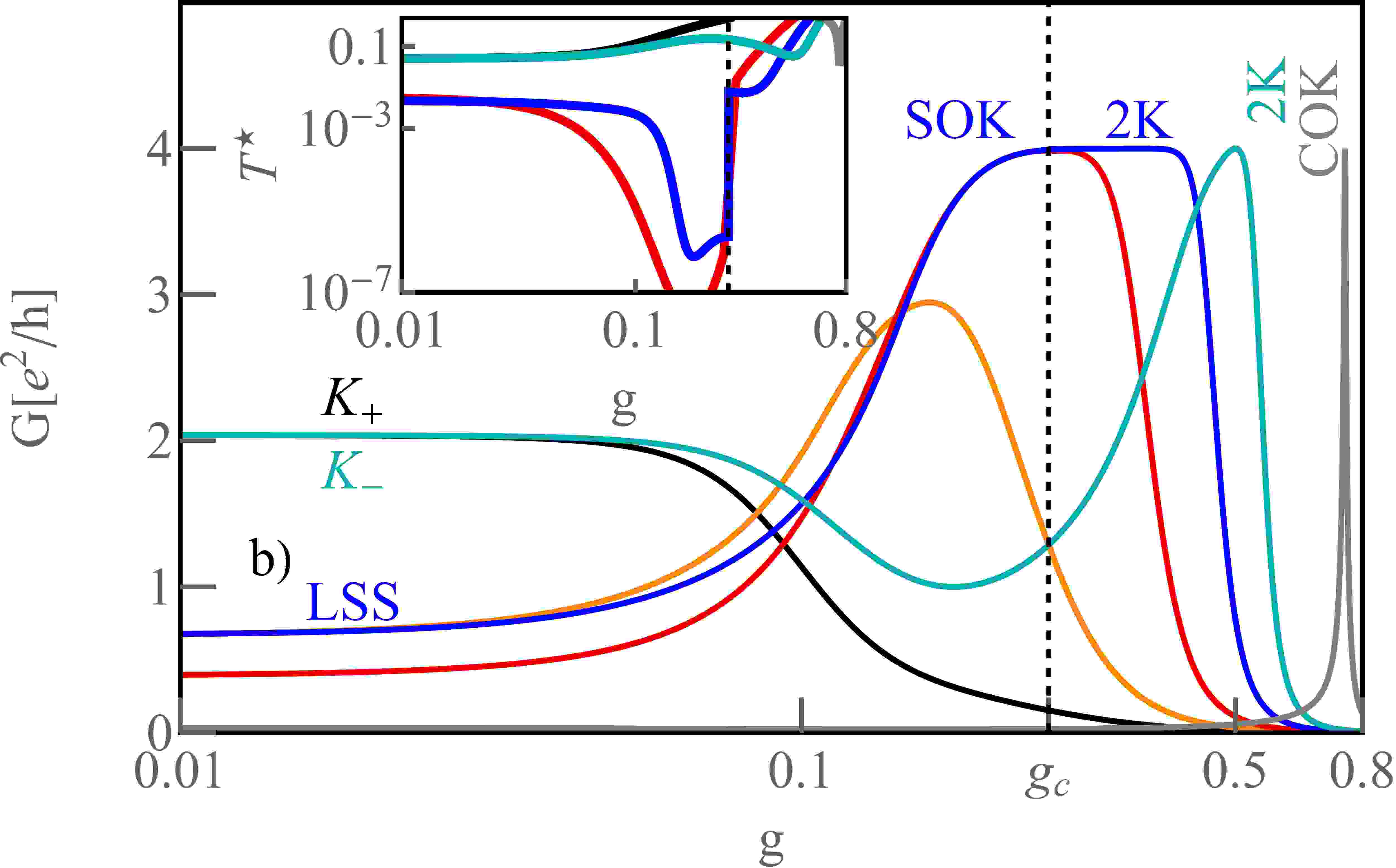}\\
\includegraphics[width=0.8\linewidth]{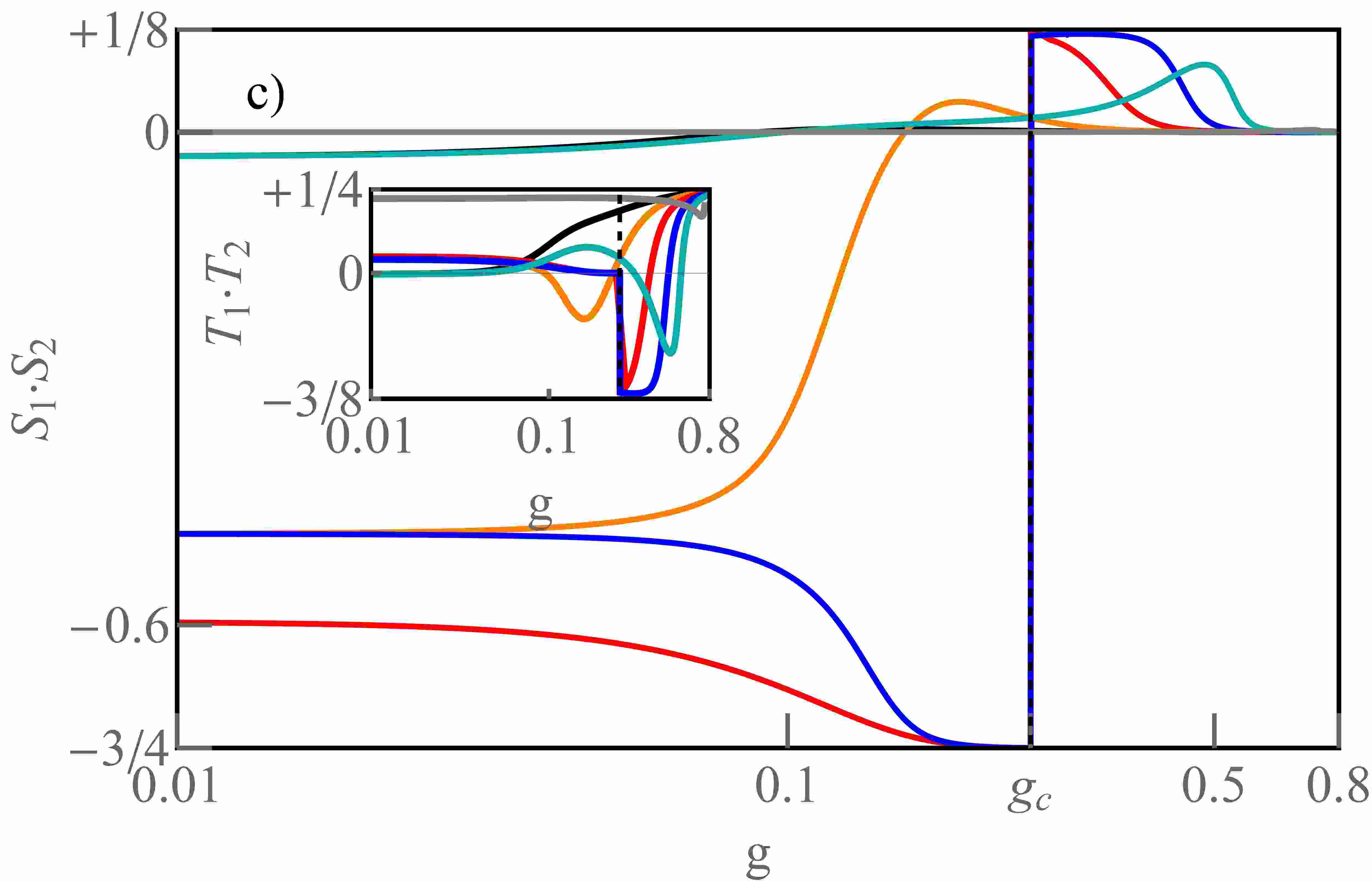}
\includegraphics[width=0.8\linewidth]{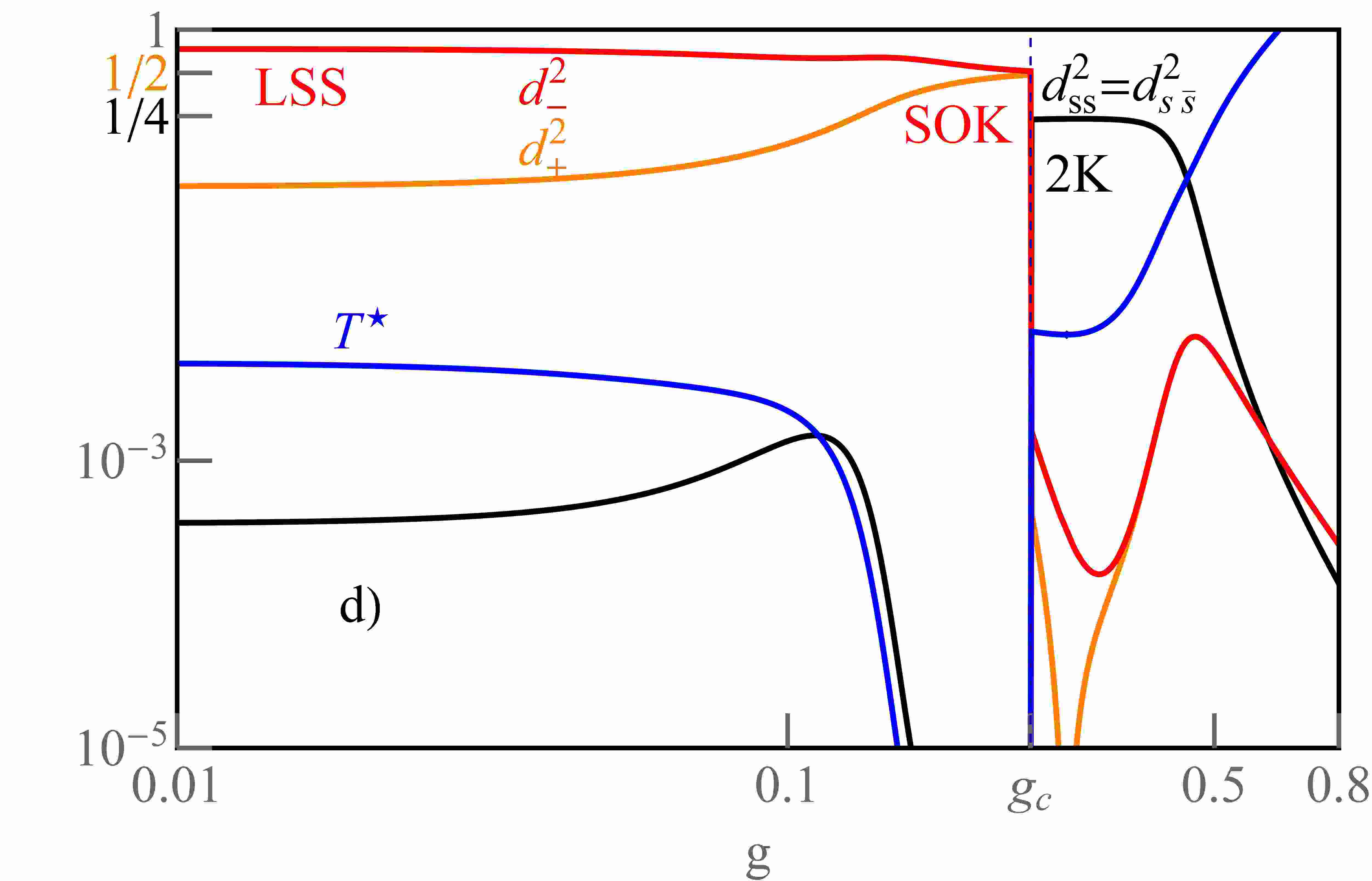}
\caption{\label{fig4} (Color online) a) Conductance map vs. dot energy level and phonon-coupling strength $g$ with marked quantum phases (see Tab.1, $g>0, U=6, U'=0, t=2, \omega_{0}=1/8$ and $\Gamma=0.1$). Dashed yellow line separates the ground sates of LSS and OK phases from the others in 2e range. Orange dot symbolizes the point $U=0$. b) Cross sections of $G$ for $E_{d}=-7, -5, -3, -1, 1, 6$ (black, orange, red, blue, cyan and gray). Inset presents the corresponding characteristic temperatures. c) Spin-spin and isospin-isospin (inset) correlators. d) Dominating salve-boson amplitudes and $T^{\star}$ for $E_{d}=-1$ presented on border of the Kondo like quantum phases SOK and 2K.}
\end{figure}
\begin{figure}
\includegraphics[width=0.8\linewidth]{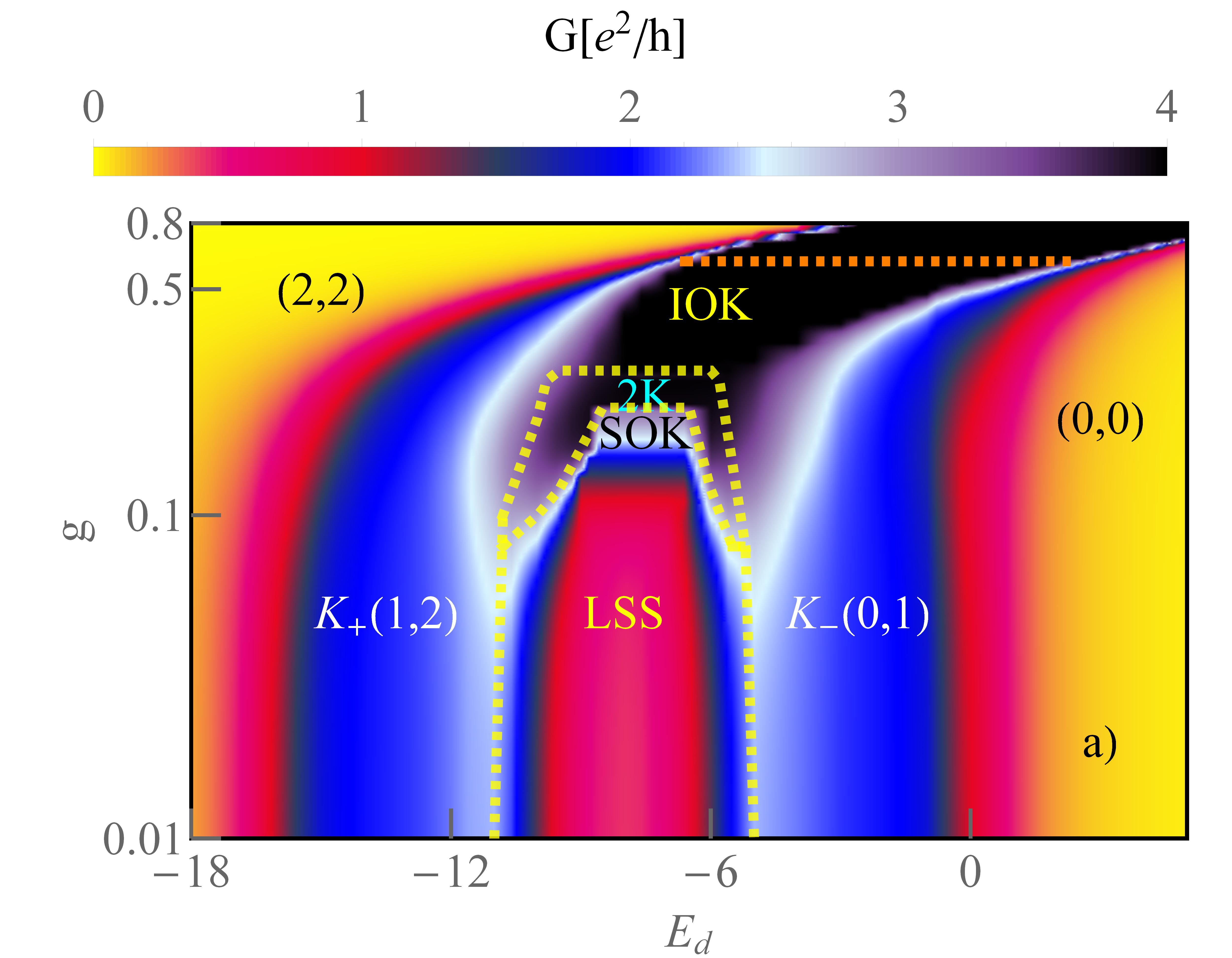}
\includegraphics[width=0.8\linewidth]{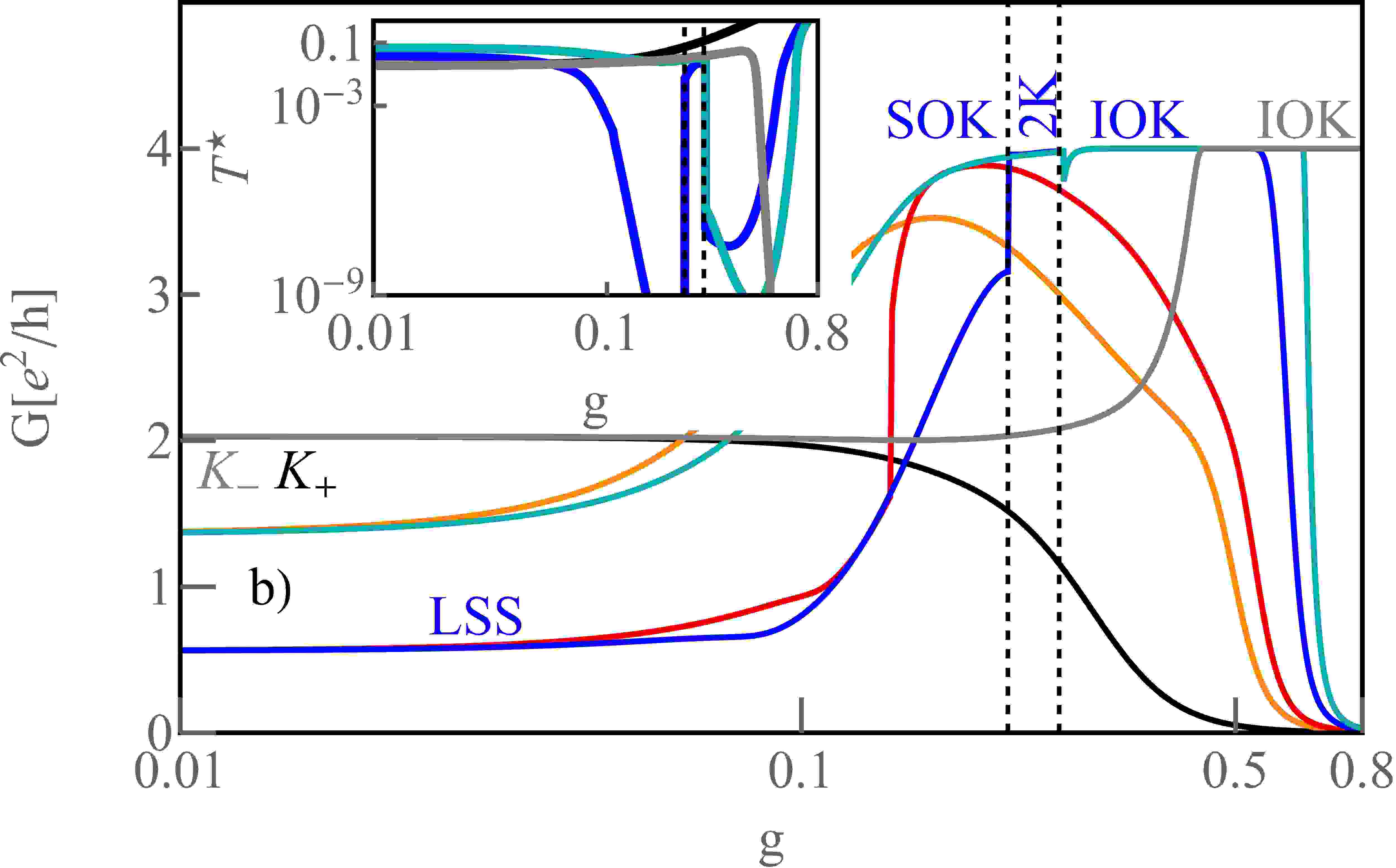}\\
\includegraphics[width=0.8\linewidth]{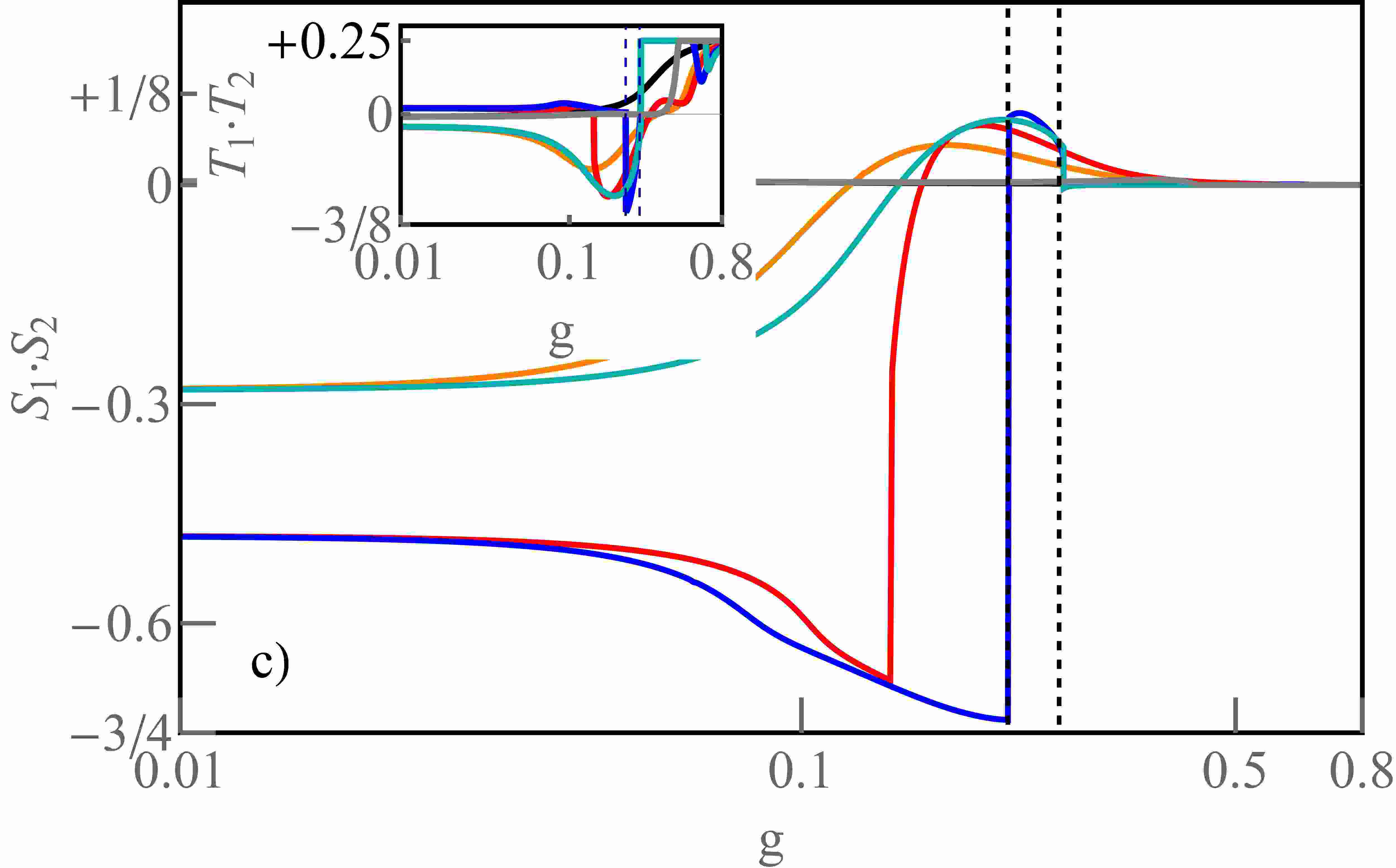}
\includegraphics[width=0.8\linewidth]{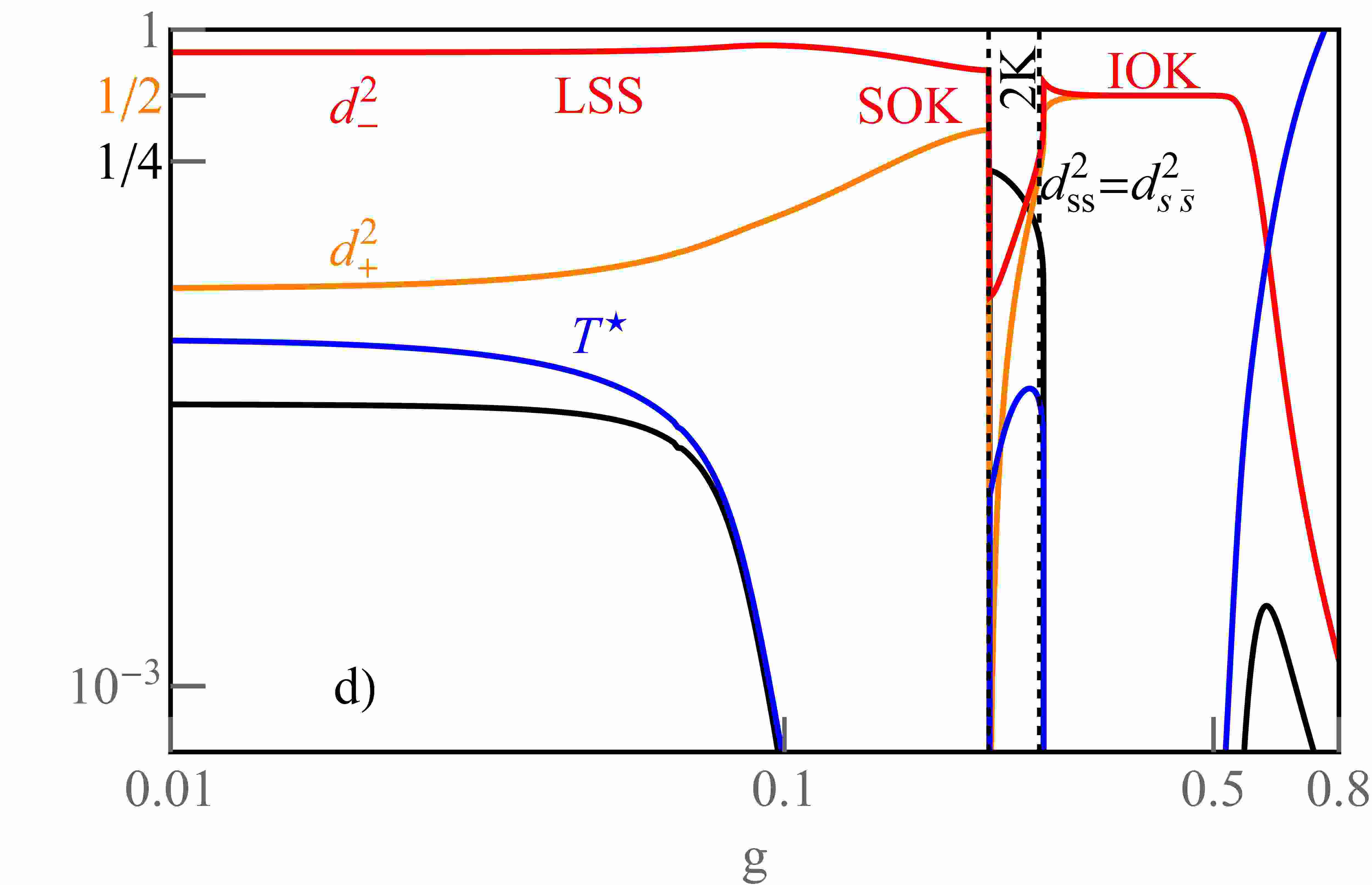}
\caption{\label{fig5} (Color online) a) Conductance map vs. dot energy level and phonon-coupling strength $g$ with marked quantum phases ($g>0, U=6, U'=5, t=0.4, \omega_{0}=1/8$ and $\Gamma=0.1$). Dashed yellow line separates the ground sates of LSS, SOK and 2K phases from the others in 2e range. Orange dashed line symbolizes the line where $U=0$.  b) Cross sections of $G$ for $E_{d}=-14, -10, -9, -7, -6, -2$ (black, orange, red, blue, cyan and gray). Inset presents the corresponding characteristic temperatures. c) Spin-spin and isospin-isospin (inset) correlators. d) Dominating salve-boson amplitudes and $T^{\star}$ for $E_{d}=-1$.}
\end{figure}

The characteristic temperatures are given by the width and the position of the quasiparticle resonances: $T^{\star}=min\{T^{\star}_{\nu}\}=\sqrt{(E_{d}\pm\widetilde{t}+\lambda_{\nu s})^{2}+(\widetilde{\Gamma} |z_{\nu s}|^{2})^{2}}$. For the case of Kondo resonance $T^{\star}=T_{K}$. In the MFA the slave boson operators are replaced by their expectation values and the stable solutions are found from the saddle point of partition function i.e. from the minimum of the free energy with respect to  the mean values of slave bosons and Lagrange multipliers. SBMFA best describes spin and orbital fluctuations in the unitary Kondo regime. For low temperatures $T<T_{K}$, and low voltages $eV<kT_{K}$ MFA's neglect of bosonic field fluctuations and renormalized level fluctuations is justified, and this is the regime we limit ourselves to in this article. Due to separability of electron and phonon degrees of freedom, which results from L-F transformation,  the dot Green's functions have the form:
\begin{eqnarray}
&&G_{\nu s}(t)=-i\theta(t)(\langle d_{\nu s}(t)d^{\dagger}_{\nu s}(0)\rangle\langle X(t)X^{\dagger}(0)\rangle
\\&&\nonumber+\langle d^{\dagger}_{\nu s}(0)d_{\nu s}(t)\rangle\langle X^{\dagger}(0)X(t)\rangle)
\end{eqnarray}
with $\langle X(t)X^{\dagger}(0)\rangle=e^{-\theta(t)}$, where $\theta(t)=(g/\omega)^{2}[n_{0}(1-e^{i\omega_{0} t})+(n_{0}+1)(1-e^{-i\omega_{0} t})]$. For $T=0$ $\langle X^{\dagger}(0)X(0)\rangle$ is reduced to $\sum^{+\infty}_{n=0}L_{n}e^{i\omega_{0}t}$, where $L_{n}=(1/n!)(g/\omega_{0})^{2n}e^{-(g/\omega_{0})^{2}}$.
The Fourier transforms of the retarded electron Green's functions of the orbitals to which the phonons are attached is given by:
\begin{eqnarray}
&&G^{R}_{\nu s}=\sum_{n}L_{n}[(1-f(E-n\omega_{0}))\widetilde{G}^{R}_{\nu s}(E-n\omega_{0})
\\&&\nonumber+f(E+n\omega_{0})\widetilde{G}^{R}_{\nu s}(E+n\omega_{0})],
\end{eqnarray}
where $f(E)$ is a Fermi function and $\widetilde{G}$ are the Green's functions determined with the use of Hamiltonian (3).
The linear conductances are  given by a Landauer-type formula:
\begin{eqnarray}
&&G=G_{+}+G_{-}=(2e^{2}/h)\sum_{\nu}sin[\pi N_{\nu}]
\\&&\nonumber=(2e^{2}/h)\sum_{\nu}T_{\nu}(0).
\end{eqnarray}
Here, transmissions are defined as:
\begin{eqnarray}
&&T_{\nu}(E)=\sum^{+\infty}_{n=-\infty}\frac{\widetilde{\Gamma}|z_{\nu s}|^{2}L_{n}}{E+n\omega_{0}-(\widetilde{E}_{d}\pm\widetilde{t}+\lambda_{\nu s})+i\widetilde{\Gamma}|z_{\nu s}|^{2}}.
\end{eqnarray}
The spin-spin and isospin-isospin correlation operators expressed by fermion operators are given by \cite{Mravlje}:
\begin{eqnarray}
&&S_{1}\cdot S_{2}=(1/4)(\sum_{ss'}n_{1s}n_{2s'}-2d^{\dagger}_{1\downarrow}d_{2\downarrow}d^{\dagger}_{2\uparrow}d_{1\uparrow}
\\&&\nonumber-2d^{\dagger}_{1\uparrow}d_{2\uparrow}d^{\dagger}_{2\downarrow}d_{1\downarrow})
\end{eqnarray}
and
\begin{eqnarray}
&&T_{1}\cdot T_{2}=(1/4)(1-\sum_{ls}n_{ls}+\sum_{ss'}n_{1s}n_{2s'}+
\\&&\nonumber2d^{\dagger}_{1\downarrow}d_{2\downarrow}d^{\dagger}_{1\uparrow}d_{2\uparrow}
+2d^{\dagger}_{2\downarrow}d_{1\downarrow}d^{\dagger}_{2\uparrow}d_{1\uparrow}).
\end{eqnarray}
The orbital (charge) fluctuations read $\Delta N^{2}_{l(\nu)}=(Q_{l(\nu)}-N_{l(\nu)})^{2}$.

\section{Results and discussion}
In the following we discuss modifications of  many-body effects in parallel double dot systems caused by electron- phonon coupling.
The calculations were performed in the strong correlation regime assuming Coulomb parameter $U=6$, coupling to the leads $\Gamma=0.1$, phonon energies $\omega_{0}=1/8$ and electron - phonon coupling strength $g$ in the range $0.01-0.8$. All numerical results are presented in the relative energy units choosing $D/50$ as the unit.

As an introduction to the description of  variety of many- body states occurring in the considered system let us first discuss the reference case of the dots in the absence of phonons and concentrate on the role  of interdot hopping $t$. Fig 2a compares conductances of DQD for different values of $t$. The dotted black line for $t=0$  and $U=U'$ illustrates the well-known SU(4) Kondo case with plateaus of $2e^{2}/h$ and $4e^{2}/h$ for $n=1,3$ and $n=2$ respectively. Dotted magenta curve presents conductance for $U'=0$ and finite $t=2$ with conductance for $n=2$ strongly reduced  compared to the unitary limit. In this range Kondo resonance destruction occurs. Other curves correspond to the cases of different intra and interdot Coulomb interactions ($U=6$, $U'=5$) and finite $t$. The charge stability regions narrow and consequently plateaus for $n=1$ and $n=3$ are replaced by  functions increasing towards the e-h symmetry point. For very  small values of $t$ many-body state forms with effective  fluctuating spin and orbital pseudospin ($p^{2}_{\pm s}=1/4$), which we call $K^{SU(4)}_{1}$ and is in fact slightly broken SU(4)  Kondo state (see Table 1).  With increasing $t$  state $K^{SU(4)}_{1,3}$ transforms into SU(2) spin  Kondo states ($K_{\pm}$) on bonding ($n=1$) or antibonding ($n=3$) orbitals respectively  with only slight change in conductance,  but drastic change in Kondo temperature (inset of Fig.3a). For $n=2$ the effect of  increase of $t$ is more decisive, because this not only  makes the description in the language of molecular orbitals more appropriate for large $t$, but also introduces spin and isospin correlations. For $t=0$, or very small values of interdot hopping the spin-orbital Kondo effect (2K) occurs, where in contrast to SU(4) effect for $n=2$,  not all six states fluctuate, but only  four of them ($d^{2}_{ss'}=1/4$, Tab. 1).  Already for $t=0$ spin and isospin correlations are non-zero ($\langle S_{1}\cdot S_{2}\rangle=+1/8$ , $\langle T_{1}\cdot T_{2}\rangle=-3/8$, Fig. 3b). For the critical value of interdot hopping $t_{c}\sim1.4\cdot10^{-3}$ sharp transition into orbital (charge)  spin correlated Kondo state SOK ($\langle S_{1}\cdot S_{2}\rangle\sim-3/4$) occurs, where cotunneling induced fluctuations between two-electron states characterized by  double occupancy  of one of the dots and vacancy of the second take place $(0,2)\leftrightarrow(2,0)$  and $G\cong4$. The isospin is screened  $\langle T_{l}\rangle=0$. Increase of tunnel coupling causes the suppression of conductance. Kondo temperature is drastically reduced. For still larger interdot tunneling  Kondo resonance is destroyed and bonding state becomes favorable. Local spin singlet  (LSS) forms in this case with antiparallel spin correlations and vanishing isospin correlations. In this case, the description in the language of the local orbitals on the dots is useful. $\langle S_{1}\cdot S_{2}\rangle$ preserves a value close to $-3/4$. Charge  fluctuations associated with the local states of the dots are negligible, those associated with the molecular states are of course finite (Fig.3d, Tab.1). Increasing hopping further results in a smooth transition to orbital spin singlet state (OSS) with weaker spin correlations than in LSS state, but also negative. Both longitudinal and transverse contributions are twice reduced in comparison to LSS (Tab.1). Isospin correlator of the bonding state in OSS state $\langle T_{1}\cdot T_{2}\rangle=1/4-1/8=+1/8$ with opposite signs of longitudinal and transverse contributions.
Fig. 3b compares the dependencies of the correlation function on the value of the hopping integral for $U'=5$ with the analogous dependencies for $U=U'=6$. In the former case transition at $t_{c}$ is sharp, whereas for the latter it is smooth. Ground states below $t_{c}$ for $U'=5$ and $U=U'=6$ are double Kondo phase (2K) and enhanced Kondo temperature phase (EKT) respectively. Both these states are characterized by the same slave boson occupations ($d^{2}_{ss'}=1/4$), Tab. 1), the  same unitary conductances $G=4e^{2}/h$, but charge fluctuations and spin and isospin correlations functions are different (Tab.1  and Fig. 3b,c). Description of 2K state  in the language of effective fluctuations between molecular states is  appropriate, whereas for EKT  both local and molecular picture play similar, less relevant  role.  Fig. 3d shows gate dependencies of local charge fluctuations at the dots ($\Delta N^{2}_{l}$) and charge fluctuations on the molecular dot orbital ($\Delta N^{2}_{\nu}$) for $n=2$ and $n=1$(inset). If orbital occupancy fluctuations  $\Delta N^{2}_{\nu}$ are smaller than local occupancy fluctuations  $\Delta N^{2}_{\nu}<\Delta N^{2}_{l}$ than orbital picture is adequate. As it seen from Fig. 3b below $t_{c}$ $\langle S_{1}\cdot S_{2}\rangle>0$, and for $t>t_{c}$ $\langle S_{1}\cdot S_{2}\rangle<0$, i.e. antiparallel spin correlations develop. Fig. 3c presents longitudinal and transverse contributions to the spin and isospin correlation functions for different two-electron ground states of DQD.

To discuss the impact of phonons on the ground-state phase diagrams we choose $t=2$ for illustration of the case of electrostatically decoupled dots ($U'=0$)  and $t=0.4$ for the capacitive coupling  $U'=5$. With this choice for $g=0$ in both cases, magnetic ground state LSS occurs in double occupancy region. Fig. 4a shows conductance  vs. dot energy and e-ph coupling constant for $U'=0$. The numbers in  brackets denote occupancies of molecular orbitals, $N_{-}$ for bonding orbital and $N_{+}$ for antibonding.  Fig. 4b presents selected, representative  cross-sections of the conductance map for the  fixed unperturbed energies of the dots. i.e. dependencies of conductance on e-ph coupling strength for different gate voltages, which for $g=0$ correspond to occupations $n=0,1,2,3$ respectively. The spin Kondo state on antibonding orbital $K_{+}$ for $n=3$ with $G\cong2$ survives until relatively strong e- ph coupling strength ($g$), for still larger coupling the dots  will be fully occupied (state $(2,2)$, $n=4$). For $n=2$  and assumed value $t=2$ the slightly perturbed LSS state ($d^{2}_{-}\sim1$), characterized by low  conductance and antiferromagnetic correlations occurs for weak coupling. This state evolves with increasing $g$ towards orbital (charge) spin correlated Kondo state SOK. For $g=g_{c}$ a sharp transition into Kondo 2K state occurs with effective fluctuations between the four spin-orbitals ($d^{2}_{ss'}=1/4$, $G\cong4$).  In 2K state interdot spin correlations are positive ($\langle S_{1}\cdot S_{2}\rangle\sim+1/8$) and Kondo temperature is much higher than for SOK state. The above described transitions from LSS state  into SOK and 2K  are  phonon induced. The spin Kondo state on bonding orbital $K_{-}$  ($n=1$) is suppressed with the increase of  e-ph coupling.  The dots becomes unoccupied ($n\cong0$).  For still stronger coupling, however,  DQD becomes doubly populated ($n\cong2$)  and double Kondo state 2K is observed. It is seen from the conductance map (Fig. 4a), that some  states from region $n=1$ develop a bit differently. Increase of g results in a change of occupation $n=1\rightarrow n=2$ and a direct transition from $K_{-}$ into 2K occurs. It is realized for example for $E_{d}=0.5$. We do not draw the cross-section for this case. Similar transitions accompany the change of occupation $n=0\rightarrow n=2$ for $E_{d}<+3$. The grey curve on Fig. 4b  ($E_{d}=6$) corresponds to transition from empty state for weak and intermediate e-ph coupling into charge-orbital Kondo state (COK) ($G\cong4$), where simultaneous effective fluctuations occur between empty and fully occupied states and between molecular orbitals ($e^{2}=f^{2}=d^{2}_{\pm}=1/4$). This is again example of phonon induced Kondo effect. COK forms for $g$ so large that intrasite interaction becomes attractive ($U<0$). This favors fluctuation of pairs of electrons or holes ($e\leftrightarrow f$). For extremely large  e-ph coupling the dots are completely filled $(2,2)$ due to the phonon induced attracting potential.

Now let us look, how the ground state picture is modified in the presence of  capacitive interdot coupling (Fig.5). Qualitatively, the transitions in areas $n=1$ and $n=3$ do not differ significantly from those discussed for the case $U'=0$. For $n=2$ the LSS ground state occurring in the weak e-ph coupling range is, similarly to the earlier discussed problem for $U'=0$, first replaced  by SOK state  and for still  higher values of $g$ by 2K.  For $g>0.26$ however, the new many-body  state appears: perturbed orbital Kondo effect, which we call  isospin correlated Kondo effect (IOK, $\langle T_{1}\cdot T_{2}\rangle=+1/4$). This state can also be reached for  strong e-ph  coupling starting at $g=0$ from the regions $n=0$ or $n=1$.

\section{Summary and conclusions}
Summarizing, in the present  paper, we examined an impact of electron-phonon coupling on electron correlations in double dot system in parallel geometry with capacitive and tunnel coupling. In the spirit of  Holstein's picture, it was assumed that phonons located on a given dot interact only with electrons on the same dot. Our analysis is addressed to molecular systems, where a strong coupling of local vibrations with electrons is expected, and to suspended quantum dots based on semiconductors, which can also be strongly influenced by phonons. Phonon-induced suppression of charging energies shifts and modifies the Coulomb blockade borders and narrows Coulomb valleys. The interplay  of weakening of both  Coulomb site energies and effective dot-lead coupling together with the changes in the separations of eigenenergies of DQD, due to phonon induced  weakening of interdot hopping, affects various electronic correlations leading to the richness of emerging phenomena.  Phonons remove or restore degeneracies destroying or reviving Kondo-like states engaging spin, orbital degrees of freedom or both of them.  Competition of intersite spin or orbital correlation with local Kondo screening is also strongly influenced by vibrations. Besides various Kondo effects: spin, orbital, spin-orbital  and charge-orbital resonances, forming due to  participation of phonons, also local spin singlets with antiferromagnetic correlations appear in  DQD coupled to local phonons. When attraction caused by phonons  exceeds site Coulomb repulsion then cotunneling processes induce effective  fluctuations of pairs of electrons or holes, which together with effective orbital fluctuations  lead to charge-orbital Kondo effect.

The issues analyzed in this work concern  elastic features of spectral densities which are reflected in the linear conductance. As far as we know, all experimental works on the Kondo effect with the participation of phonons concern inelastic effects. For verifications of our predictions on phonon induced modifications of Kondo ground state, experimental analysis of linear transport  for different frequencies and different values of the e-ph coupling  strength would be necessary. Detailed transport measurements for various phonon frequencies and different e-ph coupling strength are within the reach  of the present technology both for semiconducting QDs embedded in a freestanding membrane and for carbon nanotubes \cite{Kuo,Popov,Mariani}.

\bibliography{DKSL}{}
\end{document}